\documentclass[aps,reprint,pre,groupedaddress,longbibliography]{revtex4-2}
\usepackage{graphicx}
\usepackage{chemarrow}
\usepackage{times} 
\usepackage{xcolor}
\usepackage{amsmath,amssymb,mathrsfs,amsthm}
\usepackage{bbm}
\usepackage[colorlinks=true,allcolors=blue]{hyperref}

\begin{document}

\title{A large deviation approach to superstatistics: thermodynamic duality symmetry between conjugate variables}

\author{Shaohua Guan} 
\email[]{guanphy@163.com}
\affiliation{Defense Innovation Institute, Chinese Academy of Military Science, Beijing 100071, China}
\affiliation{Intelligent Game and Decision Laboratory, Chinese Academy of Military Science, Beijing 100071, China}

\author{Qiang Chang} 

\affiliation{Defense Innovation Institute, Chinese Academy of Military Science, Beijing 100071, China}
\affiliation{Intelligent Game and Decision Laboratory, Chinese Academy of Military Science, Beijing 100071, China}

\author{Wen Yao} 
\email[]{wendy0782@126.com}
\affiliation{Defense Innovation Institute, Chinese Academy of Military Science, Beijing 100071, China}
\affiliation{Intelligent Game and Decision Laboratory, Chinese Academy of Military Science, Beijing 100071, China}

\begin{abstract}

Superstatistics generalizes Boltzmann statistics by assuming spatio-temporal fluctuations of the intensive variables. It has many applications in the analysis of experimental and simulated data. The fluctuation of the intensity variable is the key to the validity of superstatistical theory, but the law of its distribution is still unclear. In the framework of large deviation theory, we show that the fluctuation of the intensive variable of superstatistics emerges naturally from measurements in the large data limit. Combining Bayes' theorem, we demonstrate the conditional probability distribution of the intensity variable also follows the Boltzmann statistics and the conjugate variable of the intensive variable is the extensive variable, indicating a thermodynamic duality symmetry between conjugate variables in the superstatistical systems. A new thermodynamic relation between the entropy functions of conjugate variables is obtained. We utilized a simple Ising model with fluctuating temperature to verify the dual relationship between temperature and energy. Our work may contribute to the understanding of statistical physics in complex systems and Bayesian inference.
\end{abstract}

\maketitle

%\tableofcontents

\section{Introduction}

Complex dynamical systems with intricate correlations between components lead to the nonextensive property, which is tricky for the traditional thermodynamic theory \cite{abe2001nonextensive}. Superstatistics has been introduced as generalized Boltzmann statistics to study the non-equilibrium complex systems by decomposing the time-scale separated systems into several dynamics on different time scales \cite{beck2003superstatistics,beck2004superstatistics,beck2009recent}. The superposition of several statistics arises from spatio-temporal fluctuations of the intensive variable $\beta$, which can be interpreted as the fluctuation of inverse temperature in the canonical ensemble. By introducing the distribution of $\beta$, superstatistics exhibits a robust data-fitting capability and a broad spectrum of applications, including diffusive dynamics \cite{chechkin2017brownian,itto2021superstatistical}, non-equilibrium Markovian systems \cite{lubashevsky2009generalized},random matrix theory \cite{abul2006superstatistics,abul2009superstatistical}, and hydrodynamic turbulence \cite{reynolds2003superstatistical,beck2005time,beck2007statistics}. 

In the theory of superstatistics\cite{beck2003superstatistics,beck2004superstatistics}, there are two distinct scales in time or space. At small scales, localized systems are in equilibrium, following the Boltzmann distribution. However, for complex dynamical systems, at small scales, equilibrium conditions are often not met and may instead be in a non-equilibrium steady state. Recent research indicates that statistical thermodynamic structures emerge as the asymptotic behaviors in large data limits \cite{touchette2009large,cheng2018generalized,cheng2021asymptotic,lu2022emergence}. Within the framework of large deviation theory (LDT), the generalized internal energy, entropic forces, and thermodynamic relationships are well-defined based on independently and identically distributed (iid) data series measured in the system \cite{commons2021duality,yang2022statistical,qian2022statistical}. LDT offers a potential unified approach for formulating statistical thermodynamics of non-equilibrium steady-state systems. While superstatistics has been explored from many perspectives such as the maximum entropy approach  \cite{tsallis2003constructing,van2008superstatistical}, Bayesian theory \cite{sattin2006bayesian}, and information theory \cite{yamano2006thermodynamical}, LDT can expand the theoretical framework of superstatistics theory from the perspective of statistical data analysis.

A long-standing controversial topic in classical thermodynamics and superstatistics is the distribution of $\beta$ \cite{mandelbrot1989temperature,lavenda1991probability,sattin2006bayesian,davis2018temperature,sattin2018superstatistics}. For superstatistical systems with fluctuating $\beta$, the distribution of $\beta$ depends on the mechanisms of fluctuations in the system's spatio-temporal scales. $\beta$ is assumed to follow the $\chi^2$, inverse $\chi^2$ or lognormal distributions, depending on the different ways $\beta$ is affected by microscopic random variable \cite{wilk2000interpretation,beck2001dynamical,beck2005time,beck2009recent,xu2016transition,ourabah2019superstatistics,dos2020log,gravanis2020physical,davis2023kappa}. However, the choice of $\beta$ distributions in research often depends on which distribution has better fitting ability. Thus, a key question in superstatistics is how to determine the distribution of $\beta$. For canonical ensemble, the probabilities of microscopic states and energy follow Boltzmann statistics, while the energy distribution often has a variety of distribution forms due to the energy density distribution. Therefore, whether there is a similar statistical law for the $\beta$ distribution remains a question. Revealing the universal statistical laws of $\beta$ distribution is the key to expanding the superstatistical theory and its applications.

In this paper, for superstatistical system under non-equilibrium steady-state condition at small scales, we applied LDT to analyze the emergence of superstatistical distributions, statistical laws of $\beta$ distribution, and thermodynamic relationships. In the framework of LDT, this approach clearly defines the intensive variable $\beta$ without resorting to the local equilibrium assumption and the superstatistical distribution emerges as the asymptotic behavior in the large data limit. We will show that the intensive variable $\beta$  obeys the Boltzmann-like statistics and its conjugate variable is the extensive variable, thus implying a thermodynamic duality symmetry between conjugate variables. This thermodynamic duality leads to a new thermodynamic relation between entropy functions of conjugate variables. When the superstatistical system is homogeneous, this duality symmetry will be broken and the superstatistical thermodynamics reduces to classical thermodynamics. Furthermore, we numerically verified the $\beta$ distribution follows the Boltzmann-like statistics in the Ising model with fluctuating $\beta$.

\section{Background}
\subsection{\label{ldt} A brief introduction of the application of large deviation method in thermodynamics}

Considering a stochastic system with state space $\mathcal{S}$ and a prior probability $f(s)$ of state $s$, $s \in \mathcal{S}$, the measured extensive variables of this stochastic system are an array of $K$ real observables $\boldsymbol{y(s)}$. The variable in bold indicates a vector. For simplicity, we only consider the case of $K=1$ and all variables are discrete. For the independent identically distribution (iid) data series with $N$ measured samples, the empirical mean value of $y(s)$ is $\overline{y(s)}=\sum_{i=1}^{N}y(s)_i/N$.

LDT states that, for iid data series, the probability distribution of the average value of samples has an asymptotic large deviation expression.  For data series with large and finite $N\gg1$, the probability of $\overline{y(s)}$ is 
\begin{align}
\ln \Pr(\overline{y(s)} = \overline{y}) = N \eta(\overline{y}) +o(N).
\end{align}
The negative rate function $\eta$ is related to the cumulant generating function $\psi(\beta)$ by the Legendre-Fenchel transform (LFT) \cite{tyrrell1970convex},
\begin{align}
\eta(\overline{y}) &= \mathop{\min}\limits_{\beta} \{\beta \overline{y} + \psi(\beta)\}\label{tr}, \\
\psi(\beta) &= \ln\sum\limits_{\mathcal{S}} f(s) \exp(-\beta y(s))\\
&= \ln\sum\limits_{y} f(y) \exp(-\beta y).
\end{align}
In canonical ensemble,  $\overline{y}$ and $\beta$ are the mean value of internal energy and the inverse temperature in the unit of $k_{B}$. The negative large deviation rate function $\eta(\overline{y})$ is the generalized entropy function and $\psi(\beta)$ is the logarithm of the canonical partition function. In the framework of LDT, $\beta$ represents a deviation from the prior probability distribution of microstates $f(s)$ (or the probability $f(y)$ that a state at $y$ ) \cite{yang2022statistical}. In classical statistical thermodynamics, the equal probability assumption considers the prior distribution $f(s)$ to be an equal probability distribution, and then the conjugate variables, such as the inverse temperature, quantify the deviation of the posterior distribution from the equal probability distribution. 

For the discrete variable $y$, whose values range from $\{y^1,y^2,...,y^k\}$, the empirical counting frequencies is denoted as $\boldsymbol{\nu(y)}$. It is a vector containing $\{\nu(y^1),\nu(y^2),...,\nu(y^k)\}$ and $\nu(y^i)$ represents the empirical frequency of value $y^i$. For iid data series with large and finite $N\gg1$, the probability distribution of $\boldsymbol{\nu(y)}$ is
\begin{align}
\ln \Pr(\boldsymbol{\nu(y)} \in \boldsymbol{\mathrm{d}p(y)}) = -N I(\boldsymbol{p(y)}) +o(N).  
\end{align}
The entropy function $I(\boldsymbol{p(y)})$ has the form of relative entropy
\begin{align}
I(\boldsymbol{p(y)}) = \sum_{y} p(y) \log{\frac{p(y)}{f(y)}}.\label{ef}
\end{align}
$I(\boldsymbol{p(y)})$ is related to the cumulant generating function $\lambda(\boldsymbol{\mu})$ through LFT:
\begin{align}
I(\boldsymbol{p(y)}) &= \mathop{\sup}\limits_{\boldsymbol{\mu(y)}}\{-\boldsymbol{\mu(y)}\boldsymbol{p(y)} - \lambda(\boldsymbol{\mu(y)})\}\label{tr2}, \\
\lambda(\boldsymbol{\mu(y)}) &= \ln \sum\limits_y f(y) \exp(- \mu(y)).\label{tr3}
\end{align}
The asymptotic empirical counting frequency has a canonical distribution-like form in the large data limit \cite{commons2021duality},  which is
\begin{align}
p(y) = \frac{f(y) e^{-\mu(y)}}{\sum\limits_y f(y) e^{-\mu(y)}}.\label{yy}
\end{align}
According to the Gibbs conditioning principle and maximum entropy principle \cite{yang2022statistical,qian2022statistical}, $\mu(y)$ as a generalized notion of energy is corresponding to $\beta y$, which is $\beta E$ in the conventional canonical ensemble. Then the probability of $y$ under a given $\beta$ is
\begin{align}
p(y|\beta) = \frac{f(y) e^{-\beta y}}{\sum\limits_y f(y) e^{-\beta y}}.\label{yyy}
\end{align}
In summary, for a non-equilibrium steady-state system with measured iid data series, the Large Deviation Theory provides a  framework of thermodynamics that includes the Boltzmann-like distribution (Eq. \ref{yy} and Eq. \ref{yyy}), thermodynamic relationship (Eq. \ref{tr} and Eq. \ref{tr2}) and entropy function (Eq. \ref{ef}). 

\section{Results}

\subsection{\label{iid} The measurement matrix of superstatistical system }
The typical characteristic of a superstatistical system is spatio-temporal heterogeneity. In the scenario of spatial heterogeneity \cite{beck2003superstatistics}, the basic assumption is that in a small-scale local environment, the system is in equilibrium with an inverse temperature $\beta$, and the temperature fluctuates on a large scale in space. Here, we relax this assumption and assume that the local environment is in a steady state, whether it is in equilibrium or non-equilibrium. Consider a heterogeneous environment with $M$ identical individuals (as a superstatistical system), each of which interacts with its local environment and is in a steady state. $N$ independent measurements on the observable $y$ of $M$ individuals can form a measurement matrix $\mathcal{M}$. In this $M\times N$ matrix, each unit is represented by $y_{ij}$, where $i$ represents the $i$-th individual and $j$ represents the $j$-th measurement. The range of each measurement $y_{ij}$ is in $\{y^1,y^2,...,y^k\}$. As the number of measurements $N$ increases, the empirical frequency of the measured data $y_{ij}$ in the matrix asymptotically approaches the probability $p(y)$ of the variable $y$. For the $i$-th individual, the empirical frequency of each $i$-row measurement value asymptotically approaches the probability $p_i(y)$ of its state variable $y$.

In the scenario of temporal heterogeneity, there are time scale $\tau$  on which the system itself changes rapidly and time scale $T$ on which the environment changes slowly, with a clear separation between the two time scales ($T\gg \tau$) \cite{beck2005time,xu2016transition}. For an individual in a fluctuating environment, $N$ independent measurements on observable $y$ can be performed on the time scale of $\tau$ to obtain an iid data series. On the time scale of $T$, $M$ different time periods are randomly selected to perform the above measurements. Thus, a measurement matrix $\mathcal{M}$ with $M$ rows and $N$ columns can be formed. As $M$ and $N$ increase, the empirical frequency of the data in the matrix asymptotically approaches the probability $p(y)$ of the observed variable $y$.

\begin{figure}
\includegraphics[width=8.6 cm]{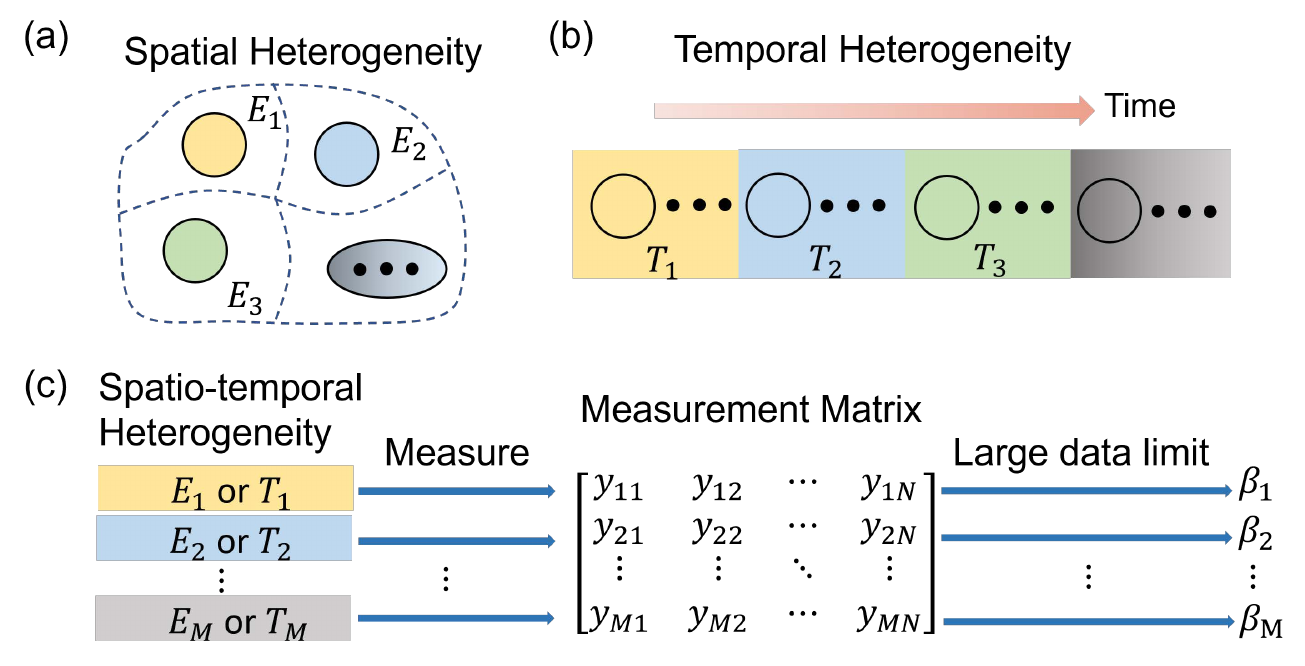}
 \caption{\label{fig1}The schematic diagram of constructing the measurement matrix in a superstatistical system with spatio-temporal heterogeneity. (a) Several identical individuals are in environments with spatial heterogeneity. (b) The single individual is in environments with temporal heterogeneity. (c) Forming a measurement matrix by measuring in different environments. In an environment with spatio-temporal heterogeneity, the iid measurement of the superstatistical system will result in a measurement matrix. In the matrix, $y_{ij}$ represents the $j$th sample of $i$th data series. Each data series has an intensive variable $\beta$ which emerges in the large data limit.}
 \end{figure}

In short, among different types of spatio-temporal heterogeneity, a measurement matrix $\mathcal{M}$ can be constructed. The probability $p(y)$ of the observed variable $y$ is the superposition of different dynamics on different spatial or time scales. The number of rows $M$ of the matrix $\mathcal{M}$ represents the number of individuals at different spatial locations in the spatial heterogeneity scenario, or represents the $M$ small time periods where the single individual is located in the temporal heterogeneity scenario. Each row of the matrix $\mathcal{M}$ contains $N$ independent and identically distributed measurements. A schematic diagram of this process is shown in Fig.~\ref{fig1}.

\subsection{\label{LDT}Superstatistics emerges from the measurement matrix}

Considering a $M\times N$ measurement matrix $\mathcal{M}$, each row with $N$ samples is an iid data series. As shown in Section \ref{ldt}, for row $i$ with large and finite $N\gg1$, the observed variable $y$ follows the Boltzmann statistics
\begin{align}
p_i(y) = \frac{f(y) e^{-\beta_i y}}{\sum\limits_y f(y) e^{-\beta_i y}}.
\end{align}
$\beta$ as the conjugate variable of $y$ emerges in the large data limit. $\beta_i$ is the value of variable $\beta$ in row $i$, which is assumed to be in the range of $\{\beta^1,\beta^2,...,\beta^l\}$. The asymptotic probability distribution of the measurement matrix is
\begin{align}
p(y)=\sum\limits_i^M p_{i}(y) / M = \sum\limits_i^M \frac{f(y) e^{-\beta_i y}}{\sum\limits_y f(y) e^{-\beta_i y}}.\label{eq_ss1}
\end{align}
$p(y)_{i}$ could be considered as conditional probability $p(y|\beta)$. Then, the summation in Eq.~(\ref{eq_ss1}) transforms to summing $\beta$,
\begin{align}
p(y)=\sum\limits_{\beta} p(\beta) &p(y|\beta)=\sum\limits_{\beta} p(\beta) \frac{f(y) e^{-\beta y}}{\mathcal{Z}_{\beta}(\beta)},\label{eq_ss2}
\end{align}
where $\mathcal{Z}_{\beta}(\beta) \equiv \sum_{y} f(y) e^{-\beta y}$ and $p(\beta)$ is the distribution of $\beta$. $f(y)$ in Eq.~(\ref{eq_ss2}) represents the prior distribution of variable $y$. By comparing Eq.~(\ref{eq_ss2}) with the theory of superstatistics, it can be found that the observed variable $y$ and the conjugate variable $\beta$ correspond to the energy and inverse temperature respectively. Therefore, Eq.~(\ref{eq_ss2}) corresponds to the type-B superstatistical distribution first proposed in Ref.~\cite{beck2003superstatistics}.

The classical theory of superstatistics assumes the system under the local equilibrium. We show that the superstatistical thermodynamic structure emerges naturally from the large data limit due to the relationship between LDT and thermodynamics. If heterogeneity in superstatistics degenerates into homogeneity, the conjugate variable $\beta$ will be a constant, and the superstatistical thermodynamics reduces to Hill's nanothermodynamics \cite{hill1994thermodynamics,lu2022emergence} and, in turn, to the classical thermodynamics in the thermodynamic limit.

\subsection{\label{beta}The statistical law of conjugate variable $\beta$}

In superstatistical theory, the conjugate variable $\beta$ has a distribution $p(\beta)$ that depends on the specific form of the spatio-temporal fluctuations. Although it is difficult to determine the exact distribution of $\beta$, it can be written as
\begin{align}
p(\beta)=\sum_y p(\beta|y)p(y).
\end{align}
$p(\beta|y)$ can be considered as conditional probability of inferring $\beta$ from a measurement of variable $y$.

\begin{figure}
\includegraphics[width=8.5 cm]{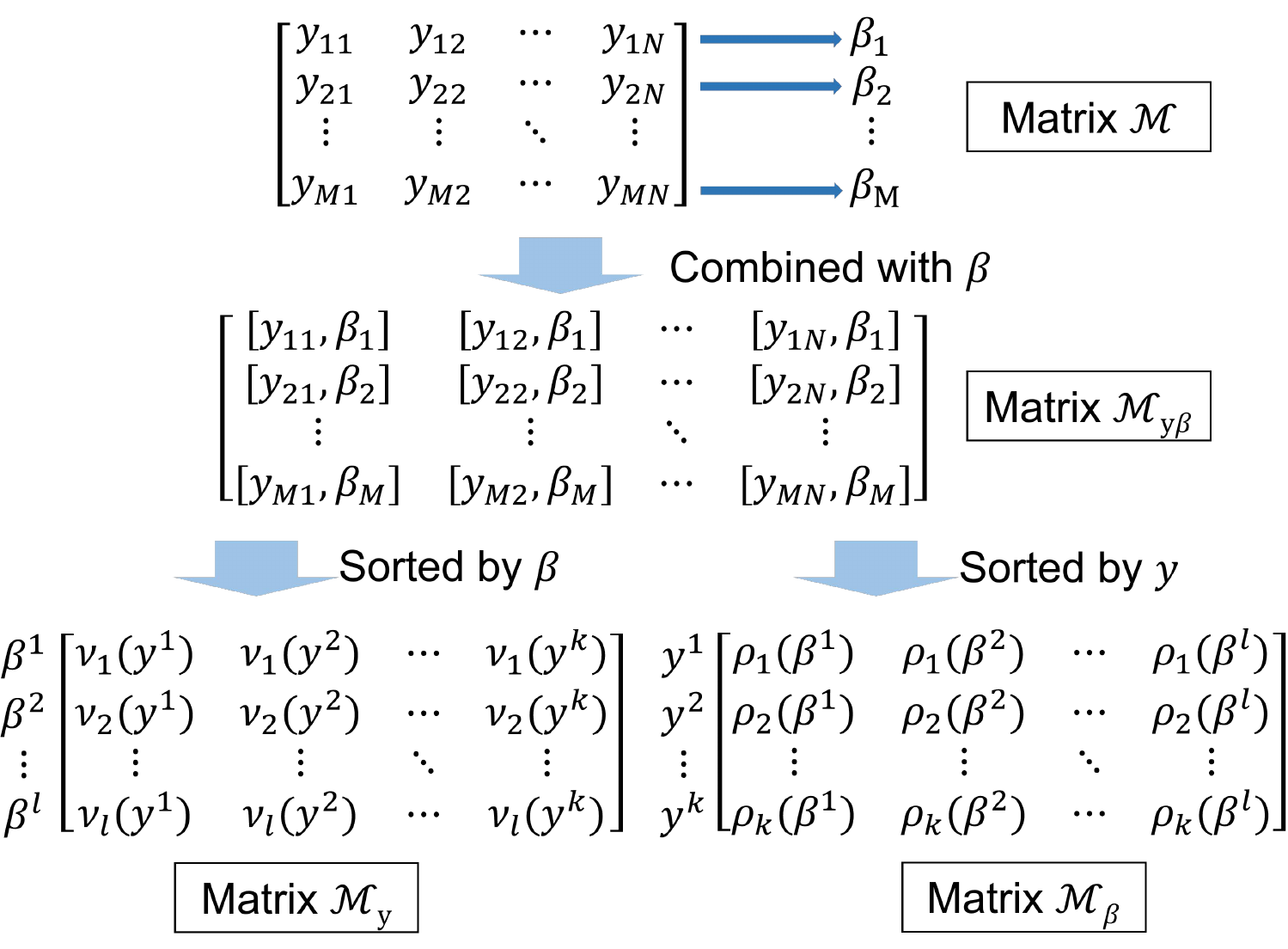}
 \caption{\label{fig2}Transformation of the measurement matrix. Each row of the measurement matrix $\mathcal{M}$ has a corresponding conjugate intensive variable $\beta$. Multiply each row of matrix $\mathcal{M}$ by its corresponding conjugate variable $\beta$ to form a new matrix $\mathcal{M}_{y\beta}$. There are $M\times N$ pairs of variables $[y,\beta]$ in the matrix $\mathcal{M}_{y\beta}$. Reorder these pairs of variables $[y,\beta]$ so that pairs with the same value $\beta^i$ or $y^i$ are placed on the same row. After re-sorting according to the value of $\beta$ or $y$, each row has the same  $\beta$ or $y$, but the amount of data in each row may be different. Extract the same $\beta$ or $y$ from each row and count the remaining empirical frequencies of another variable to form new matrices $\mathcal{M}_{y}$ and $\mathcal{M}_{\beta}$. The number of rows in the matrix $\mathcal{M}_{y}$ (or $\mathcal{M}_{\beta}$) is the number of value ranges of the variable $\beta$ (or $y$), and the number of columns is the number of value ranges of the variable $y$ (or $\beta$).}
 \end{figure}

To apply LDT to analysis the conditional probability $p(\beta|y)$, each row of the matrix $\mathcal{M}$ multiply its corresponding conjugate variable $\beta$ to form a new matrix  $\mathcal{M}_{y\beta}$ with data pairs $[y,\beta]$. $\mathcal{M}_{y\beta}$ has two variable and the probability $p(y,\beta)$ of data pairs $[y,\beta]$ has two marginal distributions $p(\beta)$ and $p(y)$. One can rearrange $\mathcal{M}_{y\beta}$ by grouping the data pairs $[y,\beta]$ with the same $\beta$ (or $y$) into the same row. Therefore, the same $\beta$ or $y$ can be extracted for each row, and the remaining data of another variable can be written as a vector of its empirical frequencies ($\boldsymbol{\nu(y)}$ or $\boldsymbol{\rho(\beta)}$). As shown in Fig.~\ref{fig2}, the two-variable matrix $\mathcal{M}_{y\beta}$ can be transformed into two matrices $\mathcal{M}_{y}$ and $\mathcal{M}_{\beta}$. According to LDT, for row $i$  in matrix $\mathcal{M}_{\beta}$ with large and finite samples, the distribution of $\beta$ of row $i$ follows the Boltzmann statistics (See Appendix \ref{AA} for detailed derivation)
\begin{align}
p_i(\beta)  = \frac{f(\beta) e^{-\alpha_i \beta}}{\sum\limits_{\beta} f(\beta) e^{-\alpha_i \beta}}.
\end{align}
 $\beta$ has its conjugate variable $\alpha$ and $\alpha_i$ is the value of variable $\alpha$. $f(\beta)$ is a prior probability of $\beta$. The choose of $f(\beta)$ determines the value of $\alpha$. The asymptotic probability distribution of $\beta$ in $\mathcal{M}_{\beta}$ is
\begin{align}
p(\beta)=\sum\limits_i^k p_i(\beta) / k = \sum\limits_i^k \frac{f(\beta) e^{-\alpha \beta}}{\sum\limits_{\beta} f(\beta) e^{-\alpha \beta}}.\label{eq_ss4}
\end{align}
If the summation in Eq.~(\ref{eq_ss4}) transforms to summing $\alpha$, then
\begin{align}
p(\beta)&=\sum\limits_{\alpha} p(\alpha) p(\beta|\alpha)=\sum\limits_{\alpha} p(\alpha) \frac{f(\beta) e^{-\alpha\beta}}{\mathcal{Z}_{\alpha}(\alpha)},\label{eqbeta}
\end{align}
where $\mathcal{Z}_{\alpha}(\alpha) \equiv \sum_{\beta} f(\beta) e^{-\alpha\beta}$. $p(y)$ in Eq.~(\ref{eq_ss2}) and $p(\beta)$ in Eq.~(\ref{eqbeta}) are the marginal distributions of the probability $p(y, \beta)$ of data pairs $[y,\beta]$ and both have superstatistical distribution forms.

% \begin{figure}
% \includegrapsics{}%
% \caption{\label{}}
% \end{figure}

\subsection{\label{analogy} The thermodynamic duality symmetry between conjugate variables}
According to the Bayes' theorem, the conditional probability of $\beta$ is
\begin{align}
p(\beta|y) = \frac{p(\beta)p(y|\beta)}{p(y)}.\label{bayes}
\end{align}
Substitute Eq. (\ref{eq_ss2}) into Eq. (\ref{bayes}), then
\begin{align}
p(\beta|y) = \frac{p(\beta)\frac{f(y)}{\mathcal{Z}_{\beta}(\beta)}\exp(- \beta y)}{p(y)}=\frac{\frac{p(\beta)}{\mathcal{Z}_{\beta}(\beta)\mathcal{C}}\exp(-y \beta)}{p(y)/f(y)\mathcal{C}}.\label{eq_ss5}
\end{align}
Eq.~(\ref{eq_ss5}) is divided by a normalized constant $\mathcal{C}\equiv \sum_{\beta}p(\beta)/ \mathcal{Z}_{\beta}(\beta)$ to ensure $p(\beta)/{\mathcal{Z}_{\beta}(\beta)\mathcal{C}}$ is a normalized probability function. Obviously, Eq.~(\ref{eqbeta}) and Eq.~(\ref{eq_ss5}) have a direct correspondence. The conjugate variable $\alpha$ is the observable $y$, $p(\beta)/ \mathcal{Z}_{\beta}(\beta) \mathcal{C}$ corresponds to the prior distribution $f(\beta)$, and $p(y)/f(y)\mathcal{C}$ is the partition function $\mathcal{Z}_{\alpha}(\alpha)$, which will be denoted as $\mathcal{Z}_{y}(y)$ later. Then Eq.~(\ref{eq_ss5}) turns to be
\begin{align}
p(\beta|y) = \frac{f(\beta)\exp(-y \beta)}{\mathcal{Z}_{y}(y)}.\label{eq_ss6}
\end{align}
Eq.~(\ref{yyy}) and Eq.~(\ref{eq_ss6}) indicate the conjugate variables $y\mathrm{\sim}\beta$ both follow the Boltzmann‐like statistics and are conjugate to each other. Moreover, Eq.~(\ref{eqbeta}) turns to be
\begin{align}
p(\beta)&=\sum\limits_{y} p(y) p(\beta|y)=\sum\limits_{y} p(y) \frac{f(\beta) e^{-y\beta}}{\mathcal{Z}_{y}(y)}.\label{eq_betadual}
\end{align}
Eq.~(\ref{eq_ss2}) and Eq.~(\ref{eq_betadual}) form a duality symmetry. The duality symmetry between conjugate variables $y$ and $\beta$ also introduces a direct analogy between statistical thermodynamics and Bayesian inference (as shown in Table \ref{tab:table1}).
\begin{table*}
\caption{\label{tab:table1}%
The correspondence between statistical thermodynamics and Bayesian inference. The left side of the table shows the energy, temperature, prior energy distribution, and canonical distribution of the canonical ensemble. The right side of the table shows the model parameters, measured variables, prior distribution, and posterior distribution in Bayesian inference. There is a direct correspondence between the canonical distribution and the Bayesian posterior distribution.}
\begin{ruledtabular}
\begin{tabular}{cccc} 
\multicolumn{2}{c}{Statistical thermodynamics}&\multicolumn{2}{c}{Bayesian inference} \\ \hline
 Energy & $y$& Model parameter & $\beta$\\ 
 Inverse temperature & $\beta$& Measured data & $y$\\ 
 Prior Energy distribution& $f(y)$&Prior probability & $f(\beta)=p(\beta)/\mathcal{Z}_{\beta}(\beta)\mathcal{C}$\\ 
  Canonical distribution &  $p(y|\beta)=f(y)\exp{(-\beta y)}/\mathcal{Z}_{\beta}(\beta)$&Posterior probability & $p(\beta|y)=f(\beta)\exp{(-y\beta) }/ \mathcal{Z}_{y}(y)$\\ 
\end{tabular}
\end{ruledtabular}
\end{table*}

The meaning of intensive parameter $\beta$ in superstatistics has been discussed from the bottom-up approach \cite{gravanis2020physical} and Bayesian approach \cite{sattin2006bayesian}. Based on the large deviation theory, we discussed the meaning of $\beta$ from the perspective of data statistics, suggesting the mathematical structure of the superstatistical dataset has a thermodynamic duality symmetry between conjugate variables. In many applications of superstatistical theory, the distribution of $\beta$ has been assumed to be several distributions, such as the $\chi^2$, inverse $\chi^2$, or lognormal distributions. The exact form of $\beta$ distribution depends on the characteristics of the system, however, we revealed that the conditional distribution $p(\beta|y)$ follows Boltzmann statistics and the distribution of $\beta$ also has an expression similar to superstatisticsal distribution. 

\subsection{\label{new_relation} Thermodynamic relations of the superstatistical system}

In classical thermodynamics, there are thermodynamic relations between free energy, entropy, intensive variables, and extensive variables independent of the underlying details. These thermodynamic quantities are well-defined and the thermodynamic relations emerge from the Gärner-Ellis theorem in the framework of LDT \cite{lu2022emergence,qian2022statistical}. Here, we will show that the thermodynamic relations in superstatistics have dual forms due to the thermodynamic duality symmetry between conjugate variables.

For the extensive variable $y$ in the row of $\mathcal{M}_{y}$ with fixed $\beta$, the relative entropy of variable $y$ with fixed $\beta$ is
\begin{equation}
\begin{aligned}
H[y|\beta)&\equiv \sum p(y|\beta) \ln \frac{p(y|\beta)}{f(y)} \\
&= \sum p(y|\beta) \ln p(y|\beta) - \sum p(y|\beta) \ln f(y)\\
&= - S[y|\beta) - \sum_y p(y|\beta)\ln f(y).
\end{aligned}
\end{equation}
 $S[y|\beta)$ is the Shannon entropy of variable $y$ with fixed $\beta$, which is $- \sum p(y|\beta)\ln p(y|\beta)$. Combined with the thermodynamic relation of Eq. (\ref{tr2}), the above formula can be written as
\begin{equation}
\begin{aligned}
-H[y|\beta)&= S[y|\beta) + \sum_y p(y|\beta)\ln f(y) \\
&= \beta \overline{y(\beta)} + \ln \mathcal{Z}_{\beta}(\beta).\label{t1}
\end{aligned}
\end{equation}
Similarly, for variable $\beta$ in the row of $\mathcal{M}_{\beta}$ with fixed $y$, there is
\begin{equation}
\begin{aligned}
-H[\beta|y) &= S[\beta|y) + \sum_{\beta} p(\beta|y)\ln f(\beta)\\
&= y\overline{\beta(y)} + \ln \mathcal{Z}_y(y).\label{t2}
\end{aligned}
\end{equation}
$H[\beta|y)$ is the relative entropy of variable $\beta$ with fixed $y$. And $S[\beta|y)$ is the Shannon entropy of variable $\beta$ with fixed $y$, which is $- \sum p(\beta|y)\ln p(\beta|y)$. Eq.~(\ref{t1}) and Eq.~(\ref{t2}) are thermodynamic relations of conjugate variables $y\mathit{\sim}\beta$ and have a dual form.

For these conjugate variables $y\mathit{\sim}\beta$, the total Shannon entropy of data pairs $[y,\beta]$ is 
\begin{align}
S[y,\beta] = S[\beta|y] + S(y) = S[y|\beta] + S(\beta).
\end{align}
$S[\beta|y]$ and $S[y|\beta]$ are the conditional entropy functions defined by
\begin{align}
S[\beta|y] = \sum_y S[\beta|y)p(y), \\
S[y|\beta] = \sum_{\beta} S[y|\beta)p(\beta).
\end{align}
After averaging Eq.~(\ref{t1}) and Eq.~(\ref{t2}) with distributions $p(\beta)$ and $p(y)$,
\begin{align}
S[y|\beta] + \sum p(y)\ln f(y) = \mathrm{<}\beta \overline{y(\beta)}\mathit{>}_{\beta} + \mathit{<}\ln \mathcal{Z}_{\beta}(\beta) \mathit{>}_{\beta},\label{a1}\\
S[\beta|y] + \sum p(\beta)\ln f(\beta)= \mathrm{<}y\overline{\beta(y)} \mathit{>}_{y} + \mathit{<}\ln \mathcal{Z}_y(y)\mathit{>}_{y}\label{a2}.
\end{align}
In the measurement matrix of superstatistical system, $\mathrm{<}\beta \overline{y(\beta)}\mathit{>}_{\beta}$ and $\mathit{<}y \overline{\beta(y)}\mathit{>}_{y}$ are equal. Hence, we get a new relation between the entropy functions of conjugate variables $y\mathit{\sim}\beta$
\begin{align}
-H(y) + \mathit{<}\ln \mathcal{Z}_{y}(y)\mathit{>}_{y}= -H(\beta) + \mathit{<}\ln \mathcal{Z}_{\beta}(\beta)\mathit{>}_{\beta} ,
\end{align}
where $H(y)$ and $H(\beta)$ are the relative entropy functions
\begin{align}
-H(y) = S(y) + \sum_{y} p(y)\ln f(y),\\
-H(\beta) = S(\beta) + \sum_{\beta} p(\beta)\ln f(\beta).
\end{align}
Therefore, the entropy functions and the averaging logarithm partition functions of conjugate variables $y\mathit{\sim}\beta$ form a new thermodynamic relation (See Appendix \ref{AB} for detailed derivation).

\subsection{\label{simple example} A simple superstatistical system: 2D Ising model with fluctuating inverse temperature}

\begin{figure*}
\includegraphics[width=16.5 cm]{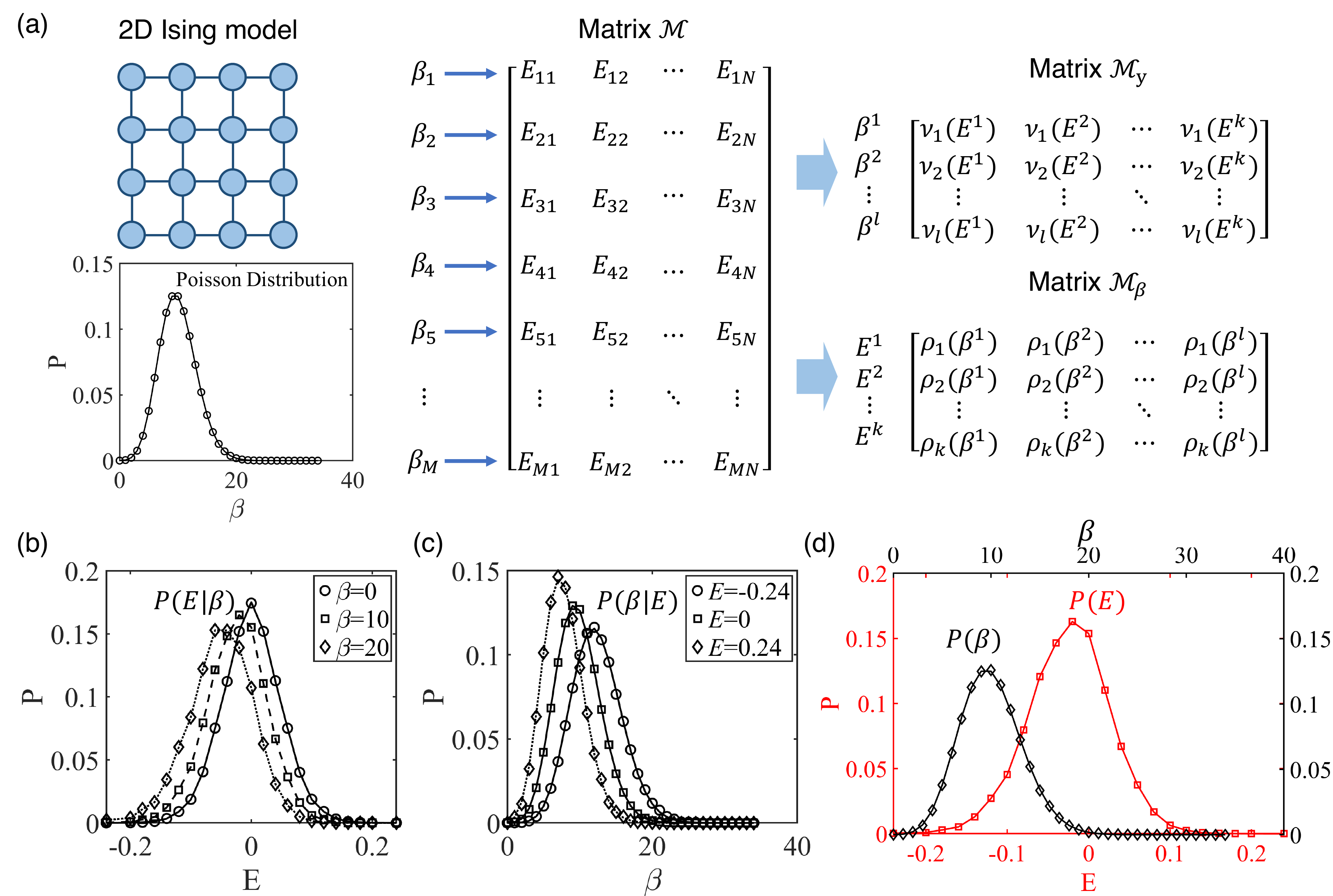}
 \caption{\label{fig3}Sampling, measurement matrix, and computation of superstatistical Ising model. (a) For a 2D Ising model with $4\times4$ spins, its inverse temperature parameter $\beta$ follows a Poisson distribution. The mean value of $\beta$ is set to $\theta=10$ and the interaction of spins is set to $J=0.01$. For simplicity, we assume that the energy unit is $1$. For the variable $\beta$, we sampled $M=10^6$ samples from the Poisson distribution. In each $\beta$, the energy state of Ising model was sampled $N=10^6$ times. The number of energy states is $k=23$. The energy state has ($k=23$) values, and the range of values for $\beta$ value is infinite. However, the probability of a Poisson distribution with a large $\beta$ is very small, so a cutoff is made for the $\beta$ value, and the value range of $\beta$ is set to $0\sim30$ ($l=31$). (b-d) Energy distributions under different environmental temperatures, $\beta$ distributions under fixed energy, and the superstatistical distribution of energy $E$ and inverse temperature $\beta$. The marked points represent the results of simulation sampling, and the line charts represent the results of analytical calculation, which are highly consistent.} 
 \end{figure*}

To verify the duality symmetry between the conjugate variables $y\mathit{\sim}\beta$, we consider a simple model which is a 2D Ising model in a zero magnetic field.  In this model, the measured variable $y$ is the energy $E$, and the conjugate variable $\beta$ is the inverse temperature. Different from the classic Ising model, the temperature fluctuates. The thermodynamics and phase transition of this type of model under two temperatures have been studied \cite{cheraghalizadeh2021superstatistical,farias2023temperature}. In order to more clearly illustrate the distribution law of $\beta$, we generalize the two-temperature Ising model to a model where the inverse temperature $\beta$ follows a Poisson distribution. The diversity of $\beta$ values leads to the complexity of calculation and simulation. To compare the simulation results with the analytical results, we select a smaller model size $4\times4$ as a compromise.

The energy of the Ising model in a zero magnetic field is 
\begin{align}
E(s)=-J\sum_{<i,j>}\sigma_i\sigma_j.
\end{align}
For a superstatistical Ising model, the energy distribution is
\begin{align}
p(E)=\sum_{\beta} p(\beta) \frac{f(E) \exp{(-\beta E)}}{\mathcal{Z}_{\beta}(\beta)} .
\end{align}
$\mathcal{Z}_{\beta}(\beta)$ is the partition function $\sum_{E}f(E)\exp{(-\beta E)}$ and $f(E)$ is the prior energy distribution of Ising model. If the distribution of $\beta$ is the Poisson distribution (the mean value of $\beta$ is denoted as $\theta$), the energy distribution is
\begin{align}
p(E)=\sum_{\beta} \frac{\theta^\beta}{\beta!}\exp{(-\theta)} \frac{f(E) \exp{(-\beta E)}}{\mathcal{Z}_{\beta}(\beta)} .
\end{align}
The conditional probability is
\begin{align}
p(\beta|E)&=\frac{p(E|\beta)p(\beta)}{p(E)}\\
&=\frac{ \frac{\theta^\beta}{\beta!}\exp{(-\theta)} \frac{f(E) \exp{(-\beta E)}}{\mathcal{Z}_{\beta}(\beta)}}{\sum_{\beta} \frac{\theta^\beta}{\beta!}\exp{(-\theta)} \frac{f(E) \exp{(-\beta E)}}{\mathcal{Z}_{\beta}(\beta)}}\\
&=\frac{\frac{\frac{\theta^\beta}{\beta!}\exp{(-\theta)}}{\mathcal{Z}_{\beta}(\beta)}\exp{(-E\beta)}}{\sum_{\beta} \frac{\theta^\beta}{\beta!}\exp{(-\theta)} \frac{ \exp{(-\beta E)}}{\mathcal{Z}_{\beta}(\beta)}}\\
&\equiv \frac{f(\beta)\exp{(-E\beta)}}{\mathcal{Z}_{E}(E)}.\label{ising_beta}
\end{align}
$f(\beta)$ as the prior probability of $\beta$ is $\frac{k^\beta}{\beta!}\exp{(-k)}/\mathcal{Z}_{\beta}(\beta)\mathcal{C}$ , $\mathcal{C}$ is the normalized constant $\sum_{\beta}\frac{k^\beta}{\beta!}\exp{(-k)}/\mathcal{Z}_{\beta}(\beta)$, and $\mathcal{Z}_{E}(E)$ as the partition function of variable $\beta$ is $\sum_{\beta} \frac{k^\beta}{\beta!}\exp{(-k)} \frac{ \exp{(-\beta E)}}{\mathcal{Z}_{\beta}(\beta)}/\mathcal{C}$. Eq. (\ref{ising_beta}) shows that the distribution of $\beta$ under fixed energy follows the Boltzmann distribution.

For the inverse temperature $\beta$ that follows a Poisson distribution, we sampled $\beta$ to obtain $M$ environments with different temperatures. For each environment, we sampled $N$ energy states of the Ising model according to the canonical distribution. Then, the measurement matrix $\mathcal{M}$, $\mathcal{M}_y$ and $\mathcal{M}_{\beta}$ are constructed. As shown in Fig. (\ref{fig3}), the simulation results of the two types of conditional distributions $p(E|\beta)$ and $p(\beta|E)$ are in agreement with the analytical results. By superimposing statistical distributions $p(E|\beta)$ and $p(\beta|E)$ under different conditions, it can obtain superstatistical distributions $p(E)$ and $p(\beta)$. By sampling and analytically calculating this simple superstatistical Ising model, we demonstrate how to construct the measurement matrix in a superstatistical system, how the conditional probability distribution of $\beta$ changes with observed quantities, and the superposition process of statistical distributions.

\section{Conclusion}

In this work, we follow the large deviation approach to investigate the thermodynamics of superstatistical systems. This work generalizes the traditional theory of superstatistics by relaxing the local equilibrium assumption to the local steady-state assumption. The superstatistical thermodynamic structure emerges from the large data limit, which is a plausible explanation for the well data-fitting ability of the superstatistical theory in many applications. The main result of our work is the thermodynamic duality symmetry between conjugate variables found in superstatistical thermodynamics. The intensive variable $\beta$ under the fixed extensive variable $y$ follows Boltzmann-like distribution, and the extensive variable $y$ acts as the conjugate variable to $\beta$. Therefore, these conjugate variables, the extensive variable, and the conjugate intensive variable, have thermodynamic duality symmetry. Moreover, the entropy functions of these conjugate variables have a new relation. This work provides a de-mechanized framework to study superstatistical thermodynamics. The duality symmetry of the conditional probabilities of conjugate variables indicates that their marginal distributions are intrinsically related. For example, the $\chi^2$ distribution of $\beta$ corresponds to the Tsallis distribution of $y$ through superstatistical theory \cite{beck2009recent}. It suggests that there are more pairs of universal distributions connected by the thermodynamic duality symmetry, which may improve our understanding of statistical distributions. 

The thermodynamic duality symmetry between conjugate variables also provides a new perspective for understanding the statistical physics of Bayesian inference. Inferring model parameters from observed data is a fundamental problem in Bayesian inference. Our results show that the distribution of model parameters can also be written as a superstatistical form, i.e., the superposition of multiple Boltzmann distributions. The important concepts in Bayesian inference (prior distribution, posterior distribution, and statistical model) have a clear correspondence in statistical thermodynamics (Table \ref{tab:table1}). However, there are some differences between superstatistical thermodynamics and Bayesian inference. The fluctuation of $\beta$ in superstatistical thermodynamics comes from the spatio-temporal heterogeneity. However, the distribution of $\beta$ in Bayesian inference represents the degree of belief. The latter is representative of our state of knowledge about the system.

\begin{acknowledgments}
All the authors designed research, performed research, and wrote the paper. S.G. thanks the members of Online Club Nanothermodynamica (Founded in June 2020). The authors declare no competing interests.
\end{acknowledgments}

\appendix
\section{Large deviation approach to the distribution of $\beta$ \label{AA}}
For the intensive variable $\beta$ in $i$th row of $\mathcal{M}_{\beta}$ with large and finite sample data (denoted as $n\gg1$), the probability of the mean value $\overline{\beta}_i$ and the empirical counting frequencies $\boldsymbol{\rho_i(\beta)}$ can be given by Cram\'{e}r's theorem and Sanov's theorem in the framework of LDT, respectively. Then
\begin{align}
\ln \Pr(\overline{\beta}_i &\in \mathrm{d}\overline{\beta}) = n \phi(\overline{\beta}) +o(n), \\
\phi(\overline{\beta}) &= \mathop{\min}\limits_{\alpha}\{\alpha\overline{\beta} + \Phi(\alpha)\}, \\
\Phi(\alpha) &= \ln\sum\limits_{\beta} f(\beta) \exp(-\alpha\beta),
\end{align}
and 
\begin{align}
\ln \Pr( \boldsymbol{\rho_i(\beta)} &\in \boldsymbol{\mathrm{d}p(\beta)}) = - n I(\boldsymbol{p(\beta)}) +o(n),  \\
I(\boldsymbol{p(\beta)}) &= \mathop{\sup}\limits_{\boldsymbol{\eta(\beta)}}\{-\boldsymbol{\eta(\beta)}\boldsymbol{p(\beta)} - \Lambda(\boldsymbol{\eta(\beta)})\},  \\
\Lambda(\boldsymbol{\eta(\beta)}) &= \ln \sum_{\beta} f(\beta) \exp(-\eta(\beta)).
\end{align}
The conjugate variable of $\beta$ is labeled by $\alpha$ and $f(\beta)$ is the reference distribution for $\beta$. In the large data limit, $\overline{\beta}$ and its conjugate variable of row $i$ converge  to $\overline{\beta}_i$ and $\alpha_i$. And the asymptotic empirical counting frequency of row $i$ converges to
\begin{align}
p_{i}(\beta) = \frac{f(\beta) e^{-\eta_{i}(\beta)}}{\sum\limits_{\beta} f(\beta) e^{-\eta_{i}(\beta)}} = \frac{f(\beta) e^{-\alpha_i \beta}}{\sum\limits_{\beta} f(\beta) e^{-\alpha_i \beta}}.
\end{align}
Then, the asymptotic probability distribution of $\beta$ in $\mathcal{M}_{\beta}$ is
\begin{align}
p(\beta)=\sum\limits_i^k p_{i}(\beta) / k = \sum\limits_i^k \frac{f(\beta) e^{-\alpha_i \beta}}{\sum\limits_{\beta} f(\beta) e^{-\alpha_i \beta}}.
\end{align}

\section{Derivation of thermodynamic relation\label{AB}}
The conditional entropy in Eq.~(\ref{a1}) and Eq.~(\ref{a2}) could be replaced by the difference between $S[y,\beta]$ and $S(\beta)$ ( and $S(y)$), then
\begin{widetext}
\begin{align}
S[y,\beta] - S[\beta] + \sum p(y)\ln f(y) &= \mathrm{<}\beta \overline{y(\beta)}\mathit{>}_{\beta} + \mathit{<}\ln \mathcal{Z}_{\beta}(\beta) \mathit{>}_{\beta},\\
S[y,\beta] - S[y] + \sum p(\beta)\ln f(\beta)&= \mathrm{<}y\overline{\beta(y)} \mathit{>}_{y} + \mathit{<}\ln \mathcal{Z}_y(y)\mathit{>}_{y}.
\end{align}
Leave the entropy $S[y,\beta]$ on the left and transfer the rest of the items to the right, then
\begin{align}
S[y,\beta]  &= S[\beta] - \sum p(y)\ln f(y) + \mathrm{<}\beta \overline{y(\beta)}\mathit{>}_{\beta} + \mathit{<}\ln \mathcal{Z}_{\beta}(\beta) \mathit{>}_{\beta},\\
S[y,\beta] &= S[y] - \sum p(\beta)\ln f(\beta) + \mathrm{<}y\overline{\beta(y)} \mathit{>}_{y} + \mathit{<}\ln \mathcal{Z}_y(y)\mathit{>}_{y}.
\end{align}
Obviously, the right side in the above two formulas is equal, then
\begin{align}
S[\beta] - \sum p(y)\ln f(y) + \mathrm{<}\beta \overline{y(\beta)}\mathit{>}_{\beta} + \mathit{<}\ln \mathcal{Z}_{\beta}(\beta) \mathit{>}_{\beta} &=
S[y] - \sum p(\beta)\ln f(\beta) + \mathrm{<}y\overline{\beta(y)} \mathit{>}_{y} + \mathit{<}\ln \mathcal{Z}_y(y)\mathit{>}_{y},\\
S[\beta] + \sum p(\beta)\ln f(\beta) + \mathrm{<}\beta \overline{y(\beta)}\mathit{>}_{\beta} + \mathit{<}\ln \mathcal{Z}_{\beta}(\beta) \mathit{>}_{\beta} &=
S[y] + \sum p(y)\ln f(y) + \mathrm{<}y\overline{\beta(y)} \mathit{>}_{y} + \mathit{<}\ln \mathcal{Z}_y(y)\mathit{>}_{y},\\
-H[\beta] +\mathit{<}\ln \mathcal{Z}_{\beta}(\beta) \mathit{>}_{\beta} &= -H[y] + \mathrm{<}y\overline{\beta(y)} \mathit{>}_{y} + \mathit{<}\ln \mathcal{Z}_y(y)\mathit{>}_{y}.
\end{align}
\end{widetext}

%\bibliographystyle{siam}
%\bibliographystyle{apsrev4-2}
%\raggedright
%\sloppy
\bibliography{TDS}

\end{document}